\documentclass[a4paper]{article}
\usepackage[latin1]{inputenc} %
\usepackage[T1]{fontenc} %
\usepackage{RR}
\usepackage{hyperref}
\usepackage{graphicx}
\usepackage{amsmath}
\usepackage{amsfonts}

\newtheorem{de}{Definition}

\newcommand{\ie}{\emph{i.e.~}}
\newcommand{\eg}{\emph{e.g.~}}
\newcommand{\cf}{\emph{cf.~}}
\newcommand{\al}{\emph{al.~}}
\newcommand{\espace}{\vspace{10pt}}

\RRdate{Juin 2009}

\RRauthor{Thomas Guyet \and Ren\'e Quiniou \and Wei Wang \and Marie-Odile Cordier}

\RRtitle{Syst\`eme auto-adaptatif de d\'etection d'intrusions Web}
\RRetitle{Self-adaptive web intrusion detection system}

\authorhead{Guyet \& al.}
\titlehead{Self-adaptive Web intrusion detection system}

\RRresume{L'\'evolution du contenu des serveurs et l'apparition de nouveaux types d'attaques rend n\'ecessaire l'adaptation dynamique des syst\`emes de d\'etection d'intrusion (IDS). De nos jours, les adaptations des IDS n\'ecessitent des interventions manuelles -- non-r\'eactives et r\'ebarbatives -- de la part des administrateurs du syst\`eme.
Dans ce papier, nous pr\'esentons un syst\`eme de d\'etection d'intrusions adaptatif qui repose sur un ensemble de diagnostiqueurs locaux. Les redondances entre les diagnostics sont exploit\'ees par un meta-diagnostiqueur qui surveille, en ligne, la consistance des diagnostics locaux, et, lorsqu'une inconsistance est d\'etect\'ee, il d\'eclenche l'adaptation des mod\`eles (des signatures d'intrusion) utilis\'es par les diagnostiqueurs incrimin\'es.
Ce syst\`eme est appliqu\'e \`a la d\'etection d'intrusions \`a partir d'un flot de requ\^etes HTTP. Les r\'esultats montrent que notre syst\`eme 1) d\'etecte les intrusions de mani\`ere pr\'ecise et sensible tout au long de la surveillance, et 2) adapte de mani\`ere ad\'equate ses mod\`eles de diagnostic et am\'eliore ainsi ses performances de d\'etection.}
\RRabstract{The evolution of the web server contents and the emergence of new kinds of intrusions make necessary the adaptation of the intrusion detection systems (IDS). Nowadays, the adaptation of the IDS requires manual -- tedious and unreactive -- actions from system administrators.
In this paper, we present a self-adaptive intrusion detection system which relies on a set of local model-based diagnosers. The redundancy of diagnoses is exploited, online, by a meta-diagnoser to check the consistency of computed partial diagnoses, and to trigger the adaptation of defective diagnoser models (or signatures) in case of inconsistency. This system is applied to the intrusion detection from a stream of HTTP requests.
Our results show that our system 1) detects intrusion occurrences sensitively and precisely, 2) accurately self-adapts diagnoser model, thus improving its detection accuracy.}

\RRmotcle{d\'etection d'intrusion, diagnostic auto-adaptation, meta-diagnostic, syst\`eme auto-adaptatif, intrusion web}
\RRkeyword{intrusion detection, self-adaptive diagnosis, meta-diagnosis, self-adaptive system, web application intrusion}

\RRprojets{DREAM}
\RRdomaine{4} 
\RRtheme{Repr\'esentation et traitement des donn\'ees et des connaissances}
\RCRennes 

\RRNo{6989} 

\begin{document}
\makeRR

\tableofcontents

\newpage
\section{Introduction}
Computerized systems, personal or professional, public or private, are more and more connected to the world wide web. Since such connections open accesses to data that the user wishes to protect, there is an increasing interest in access security issues. In order to avoid malicious accesses to, copies of or modifications of personal data, intrusion detection systems -- IDS -- are needed for robustly and precisely detecting intrusions into the protected system.

We are interested in IDSs that can detect intrusions on Web servers. Such servers are computer systems that are mostly encountered on the Web \cite{CVE2006}. Since they are accessible by everyone, not surprisingly they are among the most attacked systems. Some of the common techniques used by attackers to compromise a website include exploiting a vulnerable Web application running on the server (by attacking through improperly secured input fields), or exploiting some vulnerability present in the underlying host operating systems. In 2008 alone, Symantec identifies 12.885 site specific web application vulnerabilities. According to this recent study, 63\% of vulnerabilities affected Web applications in 2008, an increase from 59\% in 2007.
Therefore, the development of Web IDSs appears to be an important concern to respond to the need of global network security.

On the long term, the continuous accuracy and robustness of an IDS cannot be ensured in real situations without human assistance. In fact, real situations are not stable, and a major challenge is the conception of IDSs having the capacity to adapt themselves (\ie with as little human assistance as possible) to their evolving environment. The environment can evolve in three ways:
\begin{itemize}
\setlength\parskip{0in}
 \item evolution of the Web Server content: new pages and services are added and removed dynamically. Thus, requests that failed at one time might be successful later.
 \item evolution of the Web Server usage: clients modify their behavior when using the web server along the time.
 \item evolution of the intrusions: new kinds of intrusion frequently emerge.
\end{itemize}

We are particularly motivated by designing a \textbf{self-adaptive} system to detect intrusions in Web servers. Considering the evolving environment, the adaptation of the IDS is required.
Current adaptation solutions consist in updating or rebuilding some signatures (intrusion signatures or normal behavior signatures). But those solutions are tedious and unreactive because they are mostly manual. Updating a signature base requires an expert who has to perform a boring work to select the signatures that are suitable for his own server configuration and content. Similarly, rebuilding requires a tedious selection of significant trace from which the signatures are rebuilt. In both cases, a human interaction is required and the IDS will not be fully functional until it has been updated or rebuilt. In contrast, we would like to devise a \textbf{self-}adaptive Web IDS that could trigger it own adaptation without human assistance.

\espace

This situation is known as an on-line adaptation task in presence of ``concept drift'' \cite{Lane1998} and is reputed to be difficult. The design of self-adaptive IDS raises two main issues: (1) the autonomous detection and the diagnosis of the adaptation requirement, and (2) the effective adaptation of the system.

We design a multi-diagnoser architecture, associated to a meta-diagnoser, that uses integrity constraints to decide when adaptation is required and which diagnoser should adapt.
In a multi-diagnoser approach, each diagnoser agent constructs its own diagnosis from subsets of features extracted from the observations. As they are ground on different but partially redundant views of the system, these diagnoses are supposed to satisfy integrity constraints. When is not the case, \ie some of the diagnoses are contradictory, the meta-diagnoser, according to a consensus principle, detects that the observed system has evolved.

\espace

In a first section, we present the issue of the diagnosis of an evolving system. The Section \ref{sec:detectionintrusion} is dedicated to the state of the art of the intrusion detection from HTTP requests. Then, in Section \ref{sec:overview}, we give an self-adaptive diagnosis framework of the system and in Section \ref{sec:system}, we present our self-adaptive multi-diagnoser intrusion detection system. Finally, in Section \ref{sec:experiments}, we give some experiments and results.

\section{Detecting intrusions from HTTP requests}
\label{sec:detectionintrusion}

We are interested in an IDS that can detect intrusions on Web servers.
More precisely, we would like to detect intrusions from HTTP requests that are submitted to the Web server. Deployed network IDSs (\eg Snort) have the capability to detect intrusions by analyzing the traffic of TCP/IP packets. As a consequence, intruders attack the upper OSI layers, \eg the application OSI layer. Thus, IDSs dedicated to specific applications seem more relevant to tackle the complexity of these attacks from the TCP/IP packet point of view. For instance, it is easier to detect a breach in a \texttt{cgi} script from the request than from the TCP/IP traffic.
One of the main challenge is to protect the web server from known intrusions as well as unknown intrusions while avoiding any tedious update by the administrators.

In this section, firstly we introduce the structure of access logs, then we present some known intrusions using HTTP requests and finally we briefly present the general idea of HTTP request intrusion detection.

\subsection{Access logs}

\begin{figure}[ht]
 \begin{center}
  \includegraphics[width=\textwidth]{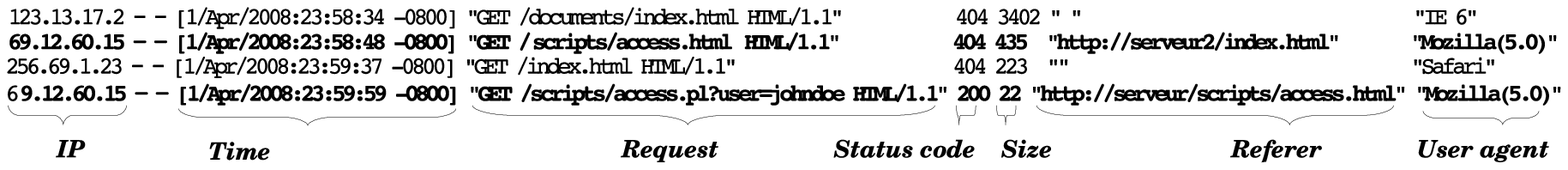}
  \caption{Apache log line examples. The session of IP 69.12.60.15 contains 2 requests.}
 \label{fig:logline}
 \end{center}
\end{figure}

All the requests received by the server are recorded in the web server access log thus a Web IDS can use the access log to detect intrusions.
A log is composed of a list of lines. Each line corresponds to a request submitted to the server and is a rich structured source of information. It has several fields that describe the request and the response made by the server to this request.
Figure \ref{fig:logline} illustrates the overall structure of a \textbf{log line} provided by an Apache server (Combined format). The main fields are:
\begin{itemize}
\setlength\parskip{0in}
\item  \textbf{IP}: the IP address of the client (remote host) who has sent the request to the server. The IP address reported here is not necessarily the address of the machine at which the user is sitting. If there exists a proxy server between the user and the server, this address will be the address of the proxy, rather than the address of the source machine.
\item \textbf{Time}: the time at which the server finished processing the request.
\item \textbf{Request}: the request line from the client is given in double quotes. The request line contains many useful pieces of information. First, it contains the method used by the client (\eg GET, POST, \dots). Second, it contains the requested resources (including the potential scripts parameters), and third, it contains the protocol used by the client (\eg HTTP/1.0).
\item \textbf{Status code}: the status code that the server sent back to the client. It indicates the kind of response the server made to the request. For example, codes beginning with 2 indicate a successful response, codes beginning with 4 indicate an error caused by the client, \dots The full list of possible status codes can be found in the HTTP specification (RFC2616 section 10).
\item \textbf{Size}: this field gives the size of the object returned to the client.
\item \textbf{Referrer}: this field gives the site that the client reports having been referred from (``-'' if not available).
\item \textbf{User agent}: The User-Agent is the identifying information that the client browser reports about itself. Especially, this field can be used to identify the robots.
\end{itemize}

\espace

A client \textbf{session} can be rebuilt from log lines by collecting the set of requests submitted by a same client (identified by his IP) to the server in a limited time window. The reconstructed session does not necessarily represent the complete client activity \cite{Murgue2006} as cache mechanisms may hide some requests. But in order to detect intrusions, the most interesting requests are those that are submitted to the server. Nonetheless, a proxy may hide an intruder behind a shared IP. We will use sessions as an alternative point of view on the current request.

\subsection{Intrusion using HTTP requests}

A Web server can be intruded by an attacker who sends a suitable HTTP request or a suitable succession of HTTP requests. Firstly, we give some simple HTTP request attacks, then we present some attacks using several requests.
\espace

Some simple attacks make use of scripts for unauthorized access to protected information such as:
\begin{itemize}
\setlength\parskip{0in}
 \item passwords (\eg \texttt{.htaccess} file),
 \item database (\eg data insertion),
 \item information about the local area network (\eg request to system commands or access to log files),
 \item information about the company activities (\eg unauthorized browse of directories).
\end{itemize}

Another common intrusion is the SQL injection where the intruder attempts to corrupt a SQL database. In such case, the intruder attempts to execute some SQL commands like \texttt{select}, \texttt{where} or \texttt{from} using the HTTP request. The same applies to attempts to execute system commands like \texttt{cat}, \texttt{grep}, \texttt{wget}, \texttt{dir}, \texttt{ls}, etc.
Specific characters like spaces, new lines or null terminators are widely used (and necessary) on most attempts to execute commands. To detect intrusion, it is interesting to look for these specific characters, but intruders often hide those characters using encoded URLs. In an encoded URL, the specific characters are encoded using hexadecimal codes, \eg '\%20' encodes a space (\cf Figure \ref{fig:IntrusionExemple}).

Figure \ref{fig:IntrusionExemple} illustrates some \texttt{awstats} attacks. We see some common system commands, separators and some encoded characters in the URL. Looking at the HTTP result code, we know that one was successfull and the other was not (error 404 and 200).

\begin{figure}
\begin{verbatim}
192.168.0.0 - - [13/Jan/2006:01:07:21 -0200] "GET /awstats/
     awstats.pl?configdir=|echo;echo%20YYY;cd%20%2ftmp%3bwget
     ...;echo%20YYY;echo|HTTP/1.0" 404 291
192.168.0.0 - - [14/Jan/2006:01:01:25 -0200] "GET /cgi-bin/
     awstats.pl?configdir=|echo;echo%20YYY;cd%20%2ftmp%3bwget
     ...;echo%20YYY;echo|HTTP/1.0" 200 291
192.168.0.1 - - [12/Apr/2006:08:05:46 -0300] "GET /rpc/..%%35%63
     ..%%35%63..%%35%63..%%35%63/winnt/system32/cmd.exe?/c+dir+
     c:\\+/OG HTTP/1.0" 400 294
192.168.0.1 - - [12/Apr/2006:08:05:47 -0300] "GET /cgi-bin/%2E
     %2E%2F%2E%2E%2F%2E%2E%%4E%4E%54%2F%73%79%73%74%65%6D%33%32
     %2Fping.exe%20127.0.0.1
192.168.0.1 - - [12/Apr/2006:08:05:43 -0300] "GET /cgi-bin/
     mrtg.cgi?cfg=/../../../../../../winnt/win.ini HTTP/1.0"
     404 289
192.168.0.2 - - [12/Apr/2006:08:05:43 -0300] "GET /cgi-bin/
     mrtg.cgi?cfg=/../../../../../../etc/passwd HTTP/1.0" 
     404 289
\end{verbatim}
\caption{Intrusion examples. Intrusions from IP 192.168.0.0 are based on the awstat script, intrusions from IP 192.168.0.1 try to illegally execute system commands and intrusions from IP 192.168.0.2 try to access to passwords.}
\label{fig:IntrusionExemple}
\end{figure}

Some complex intrusions require several HTTP requests. In their security threat report \cite{Symantec2008}, Symantec noticed increasing of complexity and sophistication of attacks. ``while a single high-security flaw can be exploited to fully compromise a user, attackers are now frequently stringing together multiple exploits for medium vulnerabilities to acheive the same goal''.
For example, a category of attacks consists in using a software security hole to install malicious software that will generate unauthorized traffic on the server.
Another example -- and the best known attack of web servers -- is the DoS attack which consists in overflooding a server with malicious requests or with requests that generate an internal error. A DoS attack provokes a system failure where the server is no more able to respond to non-intrusive requests. In such cases, the analysis of only one request may not be sufficient to detect an attack and the analysis of a session may be required.

\subsection{Intrusion Detection Systems analyzing HTTP requests}
\label{sec:IDSClasses}

The techniques for intrusion detection fall into two major categories: signature-based detection and anomaly-based detection. Signature-based detection (\eg Snort \cite{Roesch1999}, ModSecurity \cite{ModSecurity}) identifies malicious behavior by matching a behavior profile against pre-defined descriptions of attacks. Anomaly detection \cite{Denning1987}, on the other hand, defines a profile of a subject's normal behavior and attempts to identify any unacceptable deviation as the result of a potential attack.
The first category of techniques has good precision and sensibility for known intrusions but has difficulties to deal with new kinds of intrusions. In fact, a new kind of intrusion will not be detected, since its own signature is not in the intrusion signature base. On the other hand, anomaly-based techniques can easily detect unknown attacks, but their usage generates a lot of false-positive alarms.

Signature-based intrusion detection techniques are widely used for Web server intrusion detection. The request received by the Web server are successively compared with the signatures of malicious requests.
Consequently, the intrusion detection problem stands in the proposition of models (signatures) that may have the capability to represent robustly and precisely the various intrusions that may be encountered. The role of a model is, on the one hand, to focus on the request features that are relevant for intrusion detection issues and, on the other hand, to abstract the feature values into the representation.

A wide range of log line models have been already proposed.
Tombini et \al \cite{Tombini2004}'s anomaly model is a list of pairs linking the accessed ressources and the combination of parameters that were used, if any. Since web sites are organized as trees, the global model can be represented by a simple tree structure. If the requested anomaly detector belongs to the anomaly model, the anomaly detector checks whether the combination of parameters used is allowed or not.
Kruegel and Vigna \cite{Kruegel2003} introduced an anomaly-based detector of Web-based attacks. They proposed several intrusion detection models based on request features: attribute length distribution, attribute character distribution, Markov model of the structure of the query attributes, attributes order specifications, \ldots~ Ingham et \al \cite{Ingham2007a} use a deterministic automaton to model the sequence of tokens\footnote{Tokens are semantic units of the URL separated by specific characters ('\/', '?', '\&')}. Bolzoni et \al \cite{Bolzoni2008} recently proposed to use regular and irregular expressions models.
Cheng et \al \cite{Cheng2008} propose to prevent attacks by monitoring the user behavior with templates modeled by Markov models. In \cite{Singh2009, Wang2009a}, clustering techniques are used to construct dynamically a model of normal behavior as a set of clusters based on the character distribution feature.

To the best of our knowledge, no session model has been proposed so far. Nonetheless, it must be related to the recent interests in alert correlation \cite{Cuppens2002}. Several techniques, especially Bayesian Networks or causal networks, are used to combine alerts and to recognize intrusion plan \cite{Qin2004} or scenarios \cite{Ning2002}.

\subsection{Advanced IDSs}

In this paragraph, we review some specificities of advanced IDSs : 1) the combination of diagnoses and 2) the adaptation.

\subsubsection{Combining the results of intrusion detectors}

It is generally admitted that the combination of several intrusion detectors (\ie forming an ensemble of IDSs) can achieve a better performance \cite{Bass2000}. The intrusions can be better detected by combining several pieces of information that are known to be complementary. Complementary aspects can be observed along four axes:
\begin{itemize}
\setlength\parskip{0in}
 \item information that comes from distinct IDSs distributed on the local network
 \item information that is gathered from different kind of logs: different OSI layers (\eg TCP/IP packets, HTTP logs, \dots) and/or different applications, sources (system commands, database, web server, \dots).
 \item information that is extracted from the same source but through different information filters (\eg logs attributes),
 \item information that comes from systems with different ``security policies'' (\eg Signature-based vs Anomaly-based, or Anomaly detection vs Misuse detection \cite{Tombini2004}).
\end{itemize}

The first two axes use several information sources, while the last two axes combine the information obtained from only one source but from different points of view and with the aim to extract as relevant information as possible from this source.

The first two axes enable the proposition of an architecture to detect and prevent attacks in local area networks. The main idea is to centralize the information that comes from several existing tools in order to make the detection more robust.
The system of Tsian et \al \cite{Tian2005} merges alarms that comes from several network-based IDSs and host-based IDSs deployed on the local network. It uses the Dempster-Shafer \cite{Shafer1976} for data fusion. Gu et \al \cite{Gu2008} propose a decision-theoretic alert fusion based on a likelyhood ratio test (LRT).
In a global area network, a collaborative approach \cite{Michael2005, Verma2009} to intrusion detection aims at giving a global view of the network attack activity. Augmenting the information obtained at a single site with information gathered from the network can provide a more precise model of an intruder's behavior. For instance, the \textit{Worminator} \cite{Michael2005} is a P2P collaborative approach to the intrusion detection.

The last two axes are widely used in case of rich and structured data such as access logs. Since early work on web attack detection \cite{Kruegel2003}, it has been noticed that some access log line attributes are more efficient to detect some attacks than others. Similarly, some attributes generate more false alarms on some normal data.
A method based on a single attribute would be unable to detect robustly and accurately all the attacks that can be encountered on web servers, and it may be fairly easily circumvented by new attacks created by malicious clients who can hide their intrusions by avoiding the traces they know to be detectable through some monitored features.
To cover a wide range of attacks and to detect most intrusions, the Web IDS must analyze several attributes and combine the results of the analysis.
For instance, the system of Kruegel and Vigna \cite{Kruegel2003} computes an anomaly score using by a weighted sum of anomalous probabilities.

\subsubsection{Adapting an IDS}

Interaction modes between a client and a web server are highly dynamic. So, the features of normal and abnormal behaviors may change rapidly necessitating the adaptation of the monitoring system. Adapting an IDS aims at 1) progressively improving the detection reliability, and 2) at acquiring the capability to detect new kind of intrusions. We focus our attention on the discovery of new kinds of intrusion.

Practically, a Web IDS (\eg ModSecurity \cite{ModSecurity}) requires a lot of human actions, mostly tedious, which restrict the IDS reactivity. To detect new kinds of intrusion, administrators must update manually the list of intrusion patterns from signatures elaborated by experts. It appears to be strongly desirable to automatize (a part or the totality of) the discovery of new kinds of intrusion, the construction of signatures and the effective update of the IDS.

HoneyComb \cite{Kreibich2004} is a NIDS that facilitates the discovery of new kinds of intrusion by using HoneyPot. A HoneyPot is a decoy computer resource. Since there are no entry points for users to interact with these systems, activities on HoneyPots is considered suspicious by definition. Activities of entities attacking HoneyPots are logged to identify suspicious behaviors and to automatically extract intrusion signatures.

Another kind of approach aims at adapting the intrusion signatures on-line. Bojanic et \al \cite{Bojanic2005} propose to use HMM for intrusion detection in system command sequences. In this method, normal and abnormal (intrusions) behaviors are modeled by HMMs. If a sequence is suspected as being non probable with respect to known sequences, additional analyzes are performed. If these new analyzes tend to show that the sequence does not correspond to an intrusion then the HMMs linked to normal behavior are updated, else HMMs associated to intrusions are modified or a new HMM is created. In
\cite{Srinoy2006}, Srinoy proposes to use SVM intrusion models associated to a swarm intelligence technique enabling a dynamic adaptation of intrusion models.
Wang et \al \cite{Wang2009a} are confronted to the same concept drift issue as us and propose an adaptive Web intrusion detection system based on outlier detection with the affinity propagation clustering algorithm and an outlier reservoir that gathers potential intrusion waiting for further analysis.

\section{System overview}
\label{sec:overview}

\begin{figure}[ht]
 \begin{center}
  \includegraphics[width=.6\textwidth]{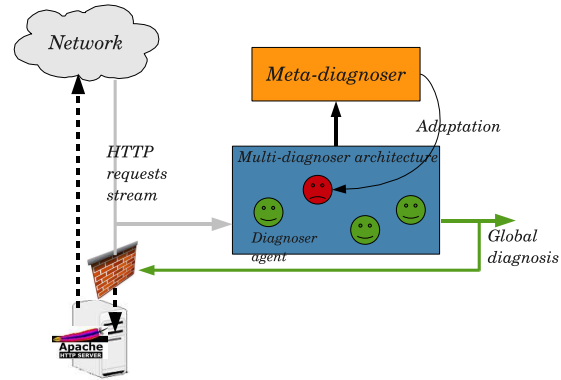}
  \caption{System overview. The multi-diagnoser architecture consists of 4 diagnosers. The decision of the red, ``unhappy'', diagnoser is inconsistent with the decisions of other diagnosers. Consequently, the meta-diagnoser may propose to adapt this diagnoser. The global diagnosis is used to block the request or not.}
 \label{fig:overview}
 \end{center}
\end{figure}

A Web server receives a stream of HTTP requests. For each new arriving request, the adaptive multi-diagnoser system constructs a diagnosis labelling the request as intrusive or not. If the request is not intrusive, then it will be processed normally by the server. In parallel a meta-diagnoser observes the diagnosis process and can trigger the adaptation of the diagnoser based on the current diagnosed request. Figure \ref{fig:overview} illustrates the system architecture.

Our diagnosis approach relies on a multi-diagnoser architecture, \ie several diagnoser agents contribute to the global diagnosis. All the diagnosers diagnose the same (sub-)problem by different methods and from different features extracted from observations taken at different time points or at different locations. As a consequence, the global diagnosis is elaborated from partial and redundant diagnosis results.

Considering that no absolute reference is available, a diagnosis mistake is detected by identifying the inconsistencies between diagnosers. The redundancy of a set of diagnoses is used by the meta-diagnoser to monitor the need for adaptation.
Our idea is to use the several partially overlapping views of the system. Due to the fact that views are partially redundant, they must be \textit{consistent} on the common parts. If not, it can be concluded that some change is occurring in some part of the system.
For example in the medical domain, at a particular time the fever severity (\texttt{high}, \texttt{average} or \texttt{low}) of a patient should be the same though it is computed by different means.
If the diagnoses are not \textit{consistent} (here, the consistency means fever severity equivalence), it is a good indication that at least one of the models should be adapted so that the diagnosers will have a better behavior in the future.

The meta-diagnoser is in charge of analyzing the output of the diagnoser agents, of detecting some \textit{inconsistencies}, of locating the diagnosers rising problems and of proposing actions to improve them by self-adaptation.

\section{An adaptive multi-diagnoser system for intrusion detection}
\label{sec:system}

As mention in Section \ref{sec:IDSClasses}, models are important to detect intrusions robustly and precisely. 
Moreover, our aim is to provide automatically the models to administrators and to avoid handmade models. To this end, we focus our attention on models that are learned from datasets of labeled log lines.

Each model focuses the diagnoser attention on a specific feature of the log line.
Several features (\eg request length, character or token distribution, status code, etc.) are extracted from observations contained in the log lines at two abstraction levels, line and session. These features are commonly used for intrusion detection \cite{Kruegel2003}. For instance, character or token distribution may be useful to distinguish malicious requests from normal ones. Some malicious requests use the URL (especially parameters to scripts) to send intrusion instructions to the target server. A perceptible change of character distribution could denote the use of such suspect instructions.

For example, to diagnose the last request of Figure \ref{fig:logline}, the features would be the following:
\begin{itemize}
\setlength\parskip{0in}
 \item character distribution of the URL in the request : \texttt{'a': 1, 'b': 0, 'c': 3, 'd': 1, 'e': 3, ...}
 \item token distribution of the URL in the request : \texttt{'scripts': 1, 'access.pl': 1, 'user': 1, 'johndoe': 1}
 \item ratio of the errorful status code in the session : \texttt{error ratio (200): 0.5}
 \item character distribution of all the URLs in the session : \texttt{'a': 2, 'b': 0, 'c': 6, 'd': 1, 'e': 4, ...}
\end{itemize}

A typical character distribution (resp. token distribution) is constructed from examples as the mean of each character (resp. token distribution) occurrences. Note that the feature space dimension is 256 for character distribution, but is infinite for token distribution. Models based on the ratio of errorful status code in sessions make use of a Gaussian distribution model and are represented by the parameters ($\sigma, \mu$) of the Gaussian.

\espace

In this section, we present a proposal for an adaptive web intrusion detection system. In Section \ref{sec:ModelBasedIDS}, we introduce the intrusion detection using model-based diagnoser agents. Then, in Section \ref{sec:MultiDiagnoserArchi}, we explain how the agents diagnoses are combined. Finally, in Section \ref{sec:AdaptativeMDS}, we detail the adaptation layer of the system.

\subsection{Model-based diagnoser to detect intrusion}
\label{sec:ModelBasedIDS}

A model-based diagnoser constructs a diagnosis about the current request according to its own model. Our definition of diagnosis is inspired by the Dempster-Shafer (DS) theory of evidence \cite{Shafer1976}. This choice is justified by the fact that the diagnosis has to take into account the uncertainty coming from the partial views that the different agents have on the observed system. To this end, the quantitative representation of a diagnosis and the explicit management of uncertainty in the DS theory are relevant. Moreover, as a fusion theory, it provides a strong solution to combine diagnoses as required by our multi-diagnoser architecture.

\subsubsection{Diagnosis with Dempster-Shafer theory of evidence}

A diagnosis expresses the more or less certainty in any of each status that can be associated to the current request: normal ($N$),  intrusive ($I$), or even unknown ($U$) \eg when the uncertainty is too high. This notion of diagnosis is formalized in the Dempster-Shafer (DS) theory of evidence.

\begin{de}
A \textbf{diagnosis} $d$ is a normalized distribution of ``masses'' on $\Omega=\left\{N,I,U\right\}$:
$$d:\Omega \mapsto [0,1], \; \sum_{A \in \Omega}d(A) = 1$$
\end{de}

For all $A\in\Omega$, $d(A)$ expresses the belief which supports the claim that the current request is of status $A$. The preferred candidate in $\Omega-\{U\}$ is called the diagnosis decision and it is unique.

\begin{de}
The \textbf{diagnosis decision}, $dd$, is the element of $\left\{N,I\right\}$ associated to the diagnosis $d$ having the maximal belief:
$$dd=\arg\max_{A \in \left\{N,I\right\}}\left(d(A)\right)$$
$\max_{A \in \left\{N,I\right\}}\left(d(A)\right)$ gives the belief in the diagnosis decision.
\end{de}

Note that the diagnosis decision cannot be $U$, but the mass $d(U)$ modifies the belief in the diagnosis decision: the greater $d(U)$, the more uncertain the diagnosis.

\subsubsection{Model-based diagnosers}

A diagnoser is characterized by its model. The model describes the diagnostic knowledge used by the diagnoser to compute its diagnosis from a subset of the observations. Each diagnoser has its proper and partial point of view on the system.

\begin{de}
A \textbf{concrete diagnoser agent} (\textbf{CDA}) is characterized by its mode $\mathcal{M}$ where $\mathcal{M}$ contains two parts: the submodel of normal requests $\mathcal{M}_N$ and the submodel of intrusive requests $\mathcal{M}_I$.
\end{de}

In a bootstrap phase, the submodels are learnt from sets of labelled examples (normal and intrusive requests). The precision $p$ of the model is computed as the ratio of correct diagnoses on the learning sets.

While diagnosing the log stream, each CDA computes its diagnosis $d$ from the distance between the current request ($R$) and the submodels: $d(N)=\|\mathcal{R} - \mathcal{M}_N\|$, $d(I)=\|\mathcal{R}-\mathcal{M}_I\|$. The uncertainty mass, $d(U)$, is given by $1-p$, where $p$ is the model precision. Finally, the diagnosis is normalized.
The distance $\| . - . \|$ depends on the request feature that is used. For character and token distributions, the model distance to a request is the euclidean distance, and for the ratio of errorful status code, it is the ratio probability given by the Gaussian distribution (\ie $\mathcal{N}_{\sigma, \mu}(ratio)$).

\subsection{Multi-diagnoser architecture}
\label{sec:MultiDiagnoserArchi}

A multi-diagnoser architecture can be seen as a multi-agent system in which the agents are diagnosers. In order to combine the diagnoses, we introduce another kind of diagnoser agents, the virtual diagnoser agent (VDA). They aim at merging the diagnosis of several other agents. As a consequence, the agents are organized in a hierarchical structure specifying the fusion scheme from the CDA diagnoses to the global diagnosis of the current request. The agents and their hierarchical structure compose the multi-diagnoser architecture.

\subsubsection{Virtual diagnoser agent}
\label{sec:VDA}

\begin{de}
A \textbf{virtual diagnoser agent} (VDA) $D^{v}$ is represented by a pair $\left<\mathcal{D}, \oplus\right>$ where $\mathcal{D}$ is the set of diagnoser agents that provide the input diagnoses to $D^{v}$ and $\oplus$ is a combination operator used by $D^{v}$ to compute its diagnosis. 
\end{de}

A VDA constructs a diagnosis by combining diagnoses that have been constructed by its related diagnoser agents (concrete or virtual) as defined in $\mathcal{D}$.
It is virtual in the sense that it is not directly related to concrete observation sources.
The combination operator $\oplus$ defines how to construct the VDA diagnosis from the diagnoses of the diagnosers of $\mathcal{D}$.
In our context, the Dempster-Shafer combination rule is used.
For all subset $A\in \Omega$, the combination of diagnoses $d_1$ and $d_2$ is computed by:
$$d(A)= (d_{1}\oplus d_{2})(A) = \frac{\sum_{B \cap C = A} d_{1}(B)d_{2}(C)}{1-\sum_{B \cap C = \emptyset} d_{1}(B)d_{2}(C)}.$$

The Dempster-Shafer combination rule is associative, thus the definition can be easily extended to the combination of more than two diagnoses. Variants of the Dempster-Shafer combination rule exists \cite{Yamada2007} and could be used as well.

\subsubsection{Multi-diagnoser architecture}
\label{sec:MDArchi}

\begin{de}
A \textbf{diagnosis combination graph} (DCG) is a directed acyclic graph where nodes represent diagnoser agents and edges specify the communication flow of diagnoses between agents. Nodes with no descendants, called \textbf{leaves}, are CDAs and other nodes are VDAs.
Among VDAs with no ancestors one is designated as the \textbf{root} node and represents the global diagnoser.
\end{de}

\begin{de}
A \textbf{multi-diagnoser architecture} is represented by a tuple $\left<\mathcal{C}, \mathcal{V}, G, R\right>$, where $\mathcal{C}$ is a set of CDAs, $\mathcal{V}$ is a set of VDAs, G is a DCG, and $R$ is the virtual diagnoser related to the root of $G$. The diagnosis computed by $R$ provides the global diagnosis of the system.
\end{de}

The diagnosis is performed recursively through the DCG: the root VDA triggers its children for monitoring. If a triggered child is a VDA, it triggers in turn new diagnosis agents; if a triggered child is a CDA, it computes a new diagnosis based on its model and the current observations. Once its diagnosis is computed, a CDA communicates its diagnosis to its ancestor (a VDA) which will combine all diagnoses sent by its children. Finally, the root VDA combines the diagnoses collected from its children and compute the global diagnosis.

\espace

\begin{figure}[tbh]
\begin{center}
 \includegraphics[width=0.7\textwidth]{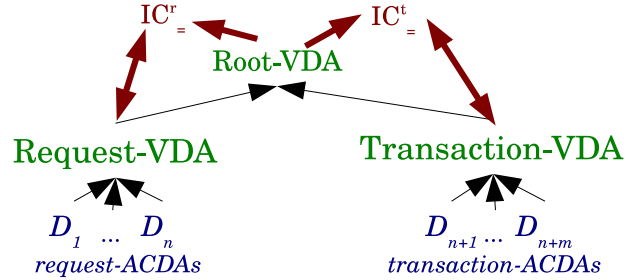}
 \caption{Diagnosis combination graph (DCG) for HTTP intrusion detection. Integrity constraints are illustrated by bold red arrows.
 }
\label{fig:HierarchieHTTP}
\end{center}
\end{figure}

The combination graph we use to detect Web intrusion (\cf Figure \ref{fig:HierarchieHTTP}) makes explicit diagnoses based on the session view and diagnoses based on the log line view: the diagnoses of the Request-CDAs (resp. the Session-CDAs) are combined by the Request-VDA (resp. the Session-VDA) and the Root-VDA combines the diagnoses of the Request-VDA and the Session-VDA.

\subsection{Adaptive multi-diagnoser system}
\label{sec:AdaptativeMDS}

\subsubsection{Integrity constraints}
\label{sec:IC}

The meta-model used for meta-diagnosis is represented by a set of integrity constraints that must be satisfied by the diagnoses computed by CDAs and VDAs. 
Integrity constraints express temporal, spatial or structural properties of the observed system.

\begin{de} An \textbf{integrity constraint} $IC$ on CDAs or VDAs $\mathcal{D} = \{D_{1}, \dots, D_{n}\}$ is a tuple $\left<\mathcal{D}, c, MDCS_{IC}\right>$ where 
\begin{itemize}
\setlength\parskip{0in}
\item $c$ is a set of constraints between the diagnoses of the diagnosers in $\mathcal{D}$,
\item $MDCS_{IC} \subset \mathcal{D}$ is the \textbf{meta-detection conflicting set}, \ie the set of the possible sources of integrity violation.
\item $D \setminus MDCS_{IC}$ is the set of \textbf{reference diagnosers}.
\end{itemize}
\end{de}

\espace

The goal is to distinguish intrusive \textit{vs} non intrusive sessions, first, and then to distinguish intrusive \textit{vs} non intrusive requests. The difficulty is then to be able to separate intrusive and non intrusive requests inside an intrusive session. To simplify the problem we assume that every requests in an intrusive session is intrusive. We are conscious that this assumption is too coarse but it could be relaxed later by supposing that a session is intrusive if it contains a high ratio of intrusive requests. Note, however, that this assumption is not directly used for computing the diagnosis but for determining whether a diagnoser should be adapted or not.
So, to design the meta-model for intrusion detection, we assume the following property for sessions: every request of a session is of the same \textit{type} as the session it belongs to, \eg if a session is intrusive ($I$), then all the requests should be intrusive. 
This assumed property is exploited by the meta-diagnoser: when a session predicts an intrusion whereas a request from the same session predicts a normal behavior an inconsistency should be reported. The converse situation may also occur.

In the context of Web IDS, the assumed property is exploited to define two integrity constraints: $IC^{r}_=$ and $IC^{t}_=$. $IC^{r}_=$ (resp. $IC^{t}_=$) is satisfied if the diagnosis decision of the Request-VDA (resp. Session-VDA) is the same as the diagnosis decision of the Root-VDA. If $IC^{r}_=$ (resp. $IC^{t}_=$) is not satisfied, the MDCS includes all the Request-CDAs (resp. Session-CDAs). Figure \ref{fig:HierarchieHTTP} represents these ICs. The red bold arrows that link an integrity constraint node to diagnosers node represent the diagnosers involved in the constraint. The red bold arrows with double arrows identifies the diagnosers of the MDCS. Diagnosers that are not in the MDCS are reference diagnosers: their diagnoses will not be contested.

\subsubsection{Meta-diagnosis}
\label{sec:MetaDiag}

A meta-detection conflicting set indicates that at least one of its elements is inconsistent with the others. The next step is to localize which diagnoser is responsible of this inconsistency and should consequently be adapted.

\begin{de}
The \textbf{meta-diagnosis set} ($MDS_{IC}$) for an integrity constraint IC is the set of agents to adapt if IC is not satisfied.
\end{de}

In our case, we assume firstly that the fusion performed by VDAs cannot be responsible of inconsistencies. Thus, the defective diagnosers must be found among the CDAs. Secondly, we consider that all the leaves that are descendants of the MDCS nodes involved in a violated integrity constraint can be suspected.
As a consequence, the meta-diagnosis set for an integrity constraint IC is the set of all the CDAs which are in $MDCS_{IC}=\{D^{c}_{1}, \dots, D^{c}_{n}\} \cup \{D^{v}_{1}, \dots, D^{v}_{_m}\}$ or which are descendants of at least one $D^{v}_{k} \in MDCS_{IC}$.

In case of inconsistency, the current global diagnosis is computed once and labelled as uncertain. This label advises the user to not trust the current diagnosis until a new trustable one, computed with adapted 
CDAs, will be provided in a near future.

\subsubsection{Adaptation}
\label{sec:Adaptation}
The final step to get a fully self-adaptive multi-diagnoser system is to have means to adapt the defective CDAs from the meta-diagnosis. To this end concrete diagnoser agents are enriched with adaptation functions.

\begin{de}
An \textbf{adaptive concrete diagnoser agent} (\textbf{ACDA}) is represented by a pair $\left<\mathcal{M}, f_{A}\right>$ where $\mathcal{M}$ is the model of a CDA and $f_A$ is a model adaptation function.
\end{de}

\begin{de}
A \textbf{reference diagnosis decision} $dd_r$ is computed by combining the diagnoses provided by the reference diagnosers of an integrity constraint IC.
\end{de}

Once, the meta diagnosis has identify the ACDAs to adapt, a \textbf{reference diagnosis decision} $dd_r$ is computed by combining the diagnoses provided by the reference diagnosers of an integrity constraint IC. For example, the integrity constraints noted $IC^r$, of the Figure \ref{fig:HierarchieHTTP}, the reference diagnoser is simply the global diagnosis (constructed by the root-VDA). In this case, there is only one reference diagnoser, then it is not require to combine several diagnoses.

The reference diagnosis decision $dd_r$ related to an unsatisfied IC is provided to the ACDAs in the $MDCS_{IC}$ for adapting their model. Continuing the previous example, diagnosis decision is provided to the Request-ACDAs (the request-VDA can not be adapted). Each agent uses it own adaptation functions with the current request and $dd_r$ as parameters. Pratically, if $dd_r=I$ (resp. $dd_r=N$), then the revised submodel $\mathcal{M}_{I}$ (resp. $\mathcal{M}_{N}$) is computed by a weighted averaging of the observed request feature (character distribution, token distribution, ...) and the old model $\mathcal{M}_{I}$ (resp. $\mathcal{M}_{N}$).

\begin{de}
An \textbf{adaptive multi-diagnoser system} is a pair $\left<\mathcal{D}, \mathcal{M_D}\right>$, where $\mathcal{D}$ is a multi-diagnoser architecture whose ACDAs are adaptive and $\mathcal{M_D}$ is a meta-model of $\mathcal{D}$. 
\end{de}

In our case, the meta-model $\mathcal{M_D}$ is a set of integrity constraints.

\subsection{Example}
\label{sec:IDS_Example}

\begin{figure}[tbh]
\begin{center}
 \includegraphics[width=0.7\textwidth]{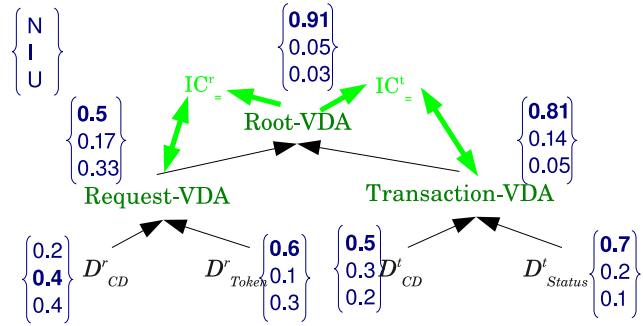}
 \caption{Diagnoses illustration (without adaptation). For each diagnosers, the 3d vector gives the diagnosis (masses distribution of N, I and U). The bold number is the highest belief and its position in the vector indicates the diagnosis decision.}
\label{fig:IDS_Example}
\end{center}
\end{figure}

Figure \ref{fig:IDS_Example} illustrates the propagation of diagnoses along the DCG. The process begins by the computing of the concrete diagnosers of the ACDAs, the results of which are given at the bottom of the figure. The diagnosis of the VDAs are computed next by combining the suitable diagnoses applying the Dempster-Shafer rule. In this example, the two ICs are satisfied because the diagnosis decision of the Root-VDA and the Request-VDA or the Session-VDA are the same~: they conclude that the diagnosis decision is $N$ and, so, the Root-VDA reports that the request is normal. Note that the diagnoser $D^r_{CD}$ and $D^r_{Token}$ disagree, but there is no IC to conclude on the dysfunction of one of them. The meta-model assumes that it is quite normal to have inconsistent diagnoses at this level and the inconsistency is solved by using the Dempster-Shafer combination rule for the fusion of the contradictory diagnoses.

\begin{figure}[tbh]
\begin{center}
 \includegraphics[width=0.7\textwidth]{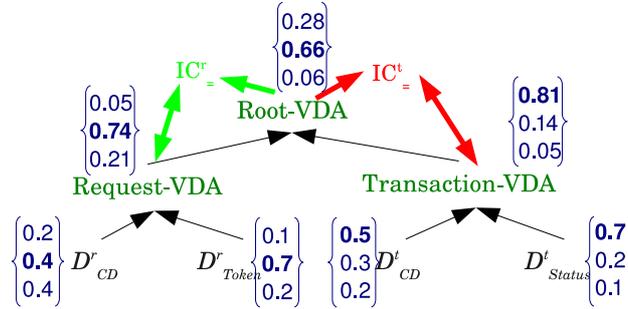}
 \caption{Diagnoses illustration (with adaptation). $IC^t_=$ is unsatisfied, $D^t_{CD}$ and $D^t_{Status}$ will be adapted.}
\label{fig:IDS_Example_2}
\end{center}
\end{figure}

Figure \ref{fig:IDS_Example_2} illustrates the case of a diagnosis which leads to an adaptation. We just changed the diagnosis of the ACDA $D^r_{Token}$. In this case, the Root-VDA diagnosis decision is $I$ but it is not equal to the Session-VDA diagnosis decision ($N$). Then, the integrity constraint $IC^t_=$ is not satisfied. The consequence will be the adaptation of the relevant submodels of ACDAs $D^t_{CD}$ and $D^t_{Status}$.

\section{Experiments}
\label{sec:experiments}

The system, called LogAnalyzer\footnote{see \textit{http://www.irisa.fr/dream/LogAnalyzer/} for more information.}, is fully implemented in C++.
The main objective of the system evaluations is to show that the multi-diagnoser approach of model adaptation improves the system performances (\ie precision and sensitivity), on the one hand, and enables the discovery and the effective use of new kinds of intrusions, on the other hand.

\subsection{Data and experiments}

We collected two large data sets of HTTP access logs on the main Apache server of two research institutes in July 2007 and June 2008 during 1 month (about 10 million of requests). A preprocessing step consists in filtering out bots and known non intrusive requests (\eg requests to static contents: \textit{.html, .jpg, .pdf}, \dots) led to a data reduction. Only 4.66\% of the original requests remained in the logs after filtering. The dataset was also checked to verify that it contained no intrusion.

For each experiment, 1 million requests, corresponding to several days of recording on our server, were extracted from the real HTTP log free of intrusion. Then, some session, 400 on average, containing 20 intrusive requests (on average) were introduced at random positions. The intrusive requests were chosen randomly among 239 known intrusive request examples from the Nikto intrusion database \cite{Nikto}.
Among them, a subset of 203 known intrusive request examples were manually selected to be used for learning the initial ACDA models. The other 36 intrusive requests were used for building instances of new kinds of intrusion that could be encountered during monitoring.

\espace

For each request, we compared the global diagnosis decision to the known status (intrusion or normal) of the request and it were classified among :
\begin{itemize}
\setlength\parskip{0in}
 \item the false positives (FP) : normal requests that have been diagnosed as intrusive,
 \item the false negatives (FN) : intrusive requests that have been diagnosed as normal,
 \item the true negatives (TN) : normal requests that have been diagnosed successfully as normal,
 \item or, the true positives (TP) : intrusive requests that have been diagnosed successfully as intrusive.
\end{itemize}

We segmented the log in 100 batches of 10.000 requests. For each batch, we counted the number of FP, TP, FN and TN occurring in the batch, and we computed the following monitoring performance indicators:
\begin{itemize}
\setlength\parskip{0in}
 \item Detection rate ($DR=TP/(TP + FN)$), \ie the accurrately recognized intrusions.
 \item False Positive rate ($FPR=FP/(FP + TP + FN + TN)$), 
 \item F-measure ($F-Measure=2*DR*P/(DR+P)$ where $P=VP/(VP+FP)$).
\end{itemize}

In this way, it is possible to observe the evolution of performance indicators over time.

In the experiments, we studied the accuracy of adaptations and the improvement of the detection performances by adaptations. The experiments consisted in comparing diagnosis performances with and without adaptation. Without adaptation, the diagnoses were computed according to the principle of our multi-diagnoser architecture but the models of the diagnoser agents, learned from the training set, do not evolve.

\espace

Experiments have been performed using a personal computer (Intel Centrino Duo T7500). It takes less than 25' to process the 1 million request and it requires less than 30 Mo of memory. The table \ref{tab:computingtime} illustrates the computing times we obtained with a personal computer (Intel Centrino Duo T7500).

\begin{table}
\center
\begin{tabular}{|p{5cm}|c|c|}
\hline
 & \textbf{Diagnosis time} & \textbf{Adaptation time}\\ \hline
Logline character distribution & 31'' & 2'' \\ \hline
Logline tokens distribution & 16'11'' & 6'10''\\ \hline
Session character distribution & 42'' & $\epsilon$\\ \hline
Session error proportion & 21'' & $\epsilon$\\ \hline
Global & 23'48''& \\ \hline
\end{tabular}
\caption{Cumulate time spend by agents (or the system) to diagnose or to adapt the 1 million requests. $epsilon$ means less than 1 second.}
\label{tab:computingtime}
\end{table}

\subsection{Results}

\begin{figure}[tbh]
\begin{center}
 \includegraphics[width=0.49\textwidth]{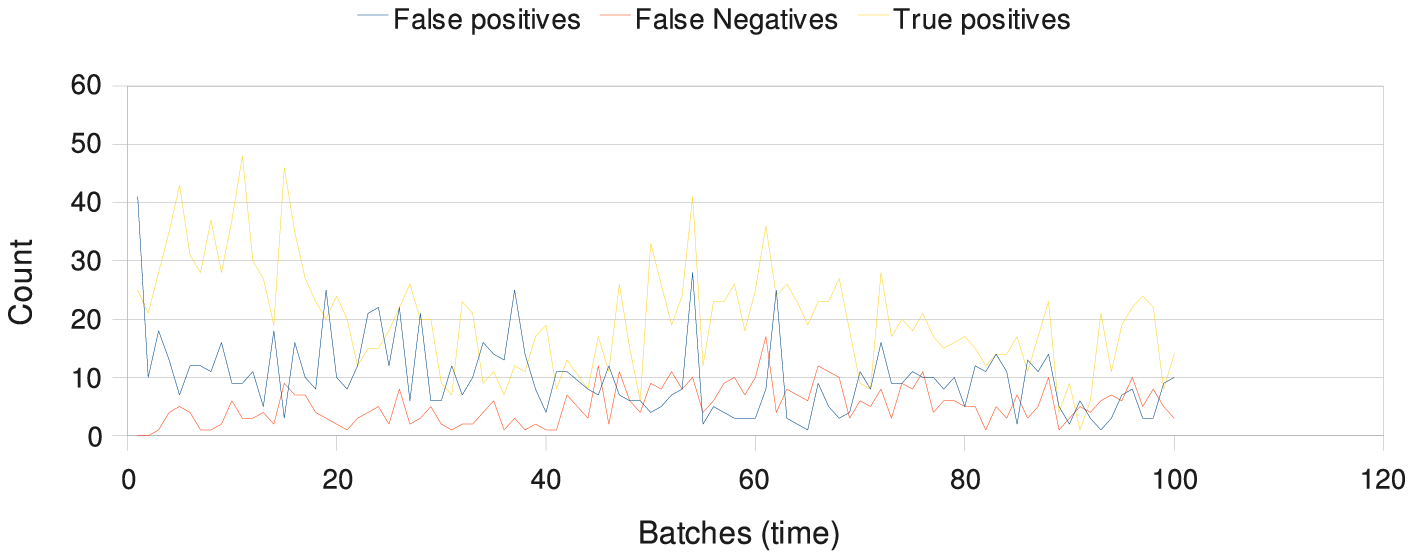}
\includegraphics[width=0.49\textwidth]{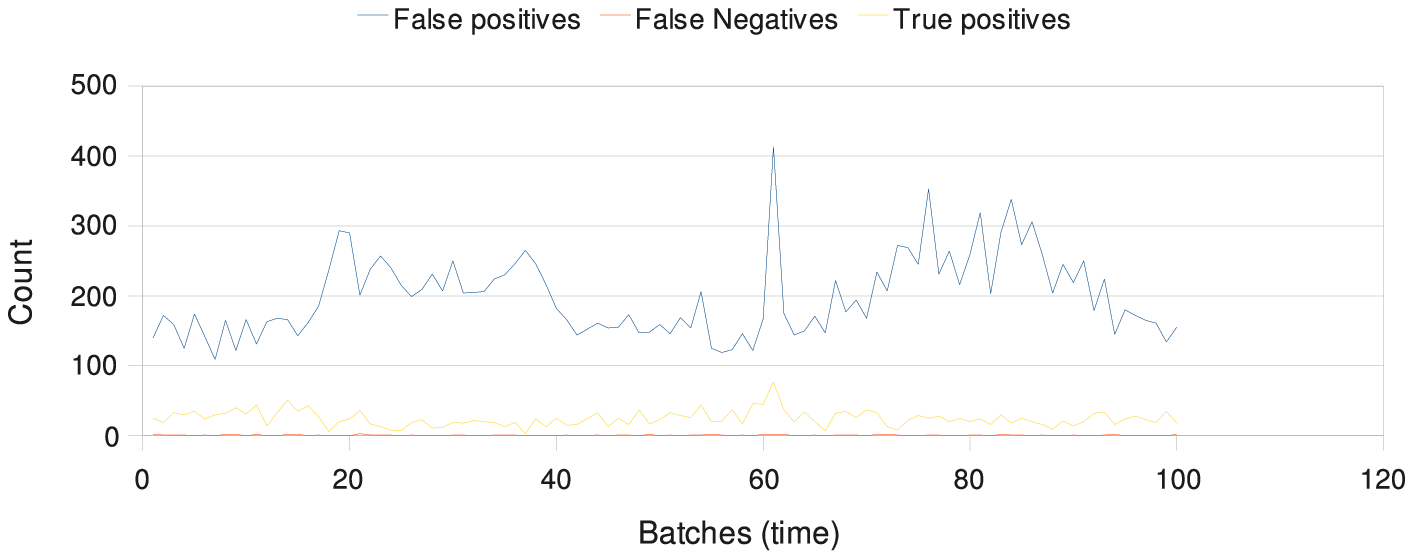} \\
 \includegraphics[width=0.49\textwidth]{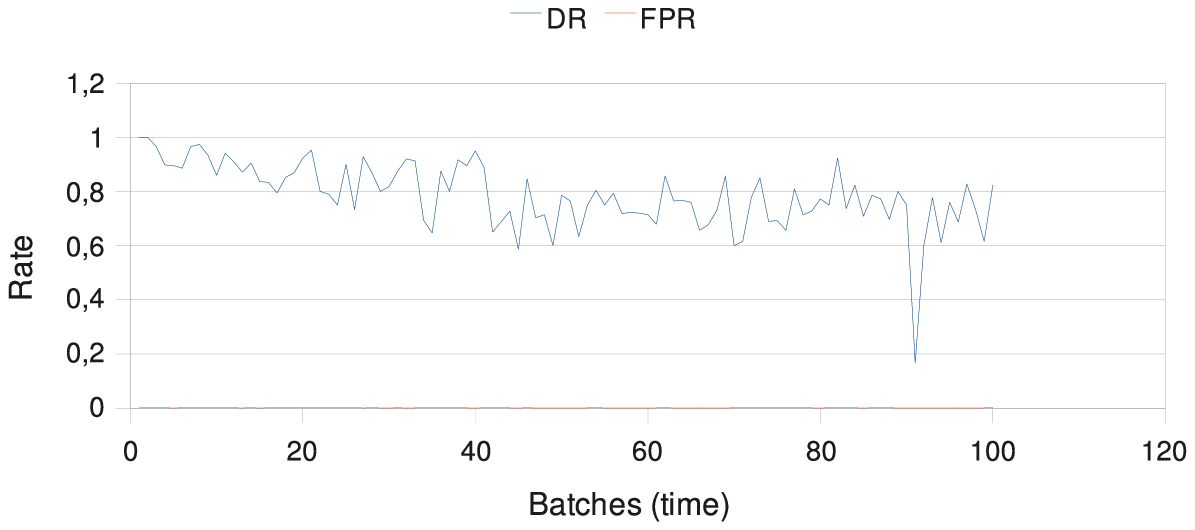}
\includegraphics[width=0.49\textwidth]{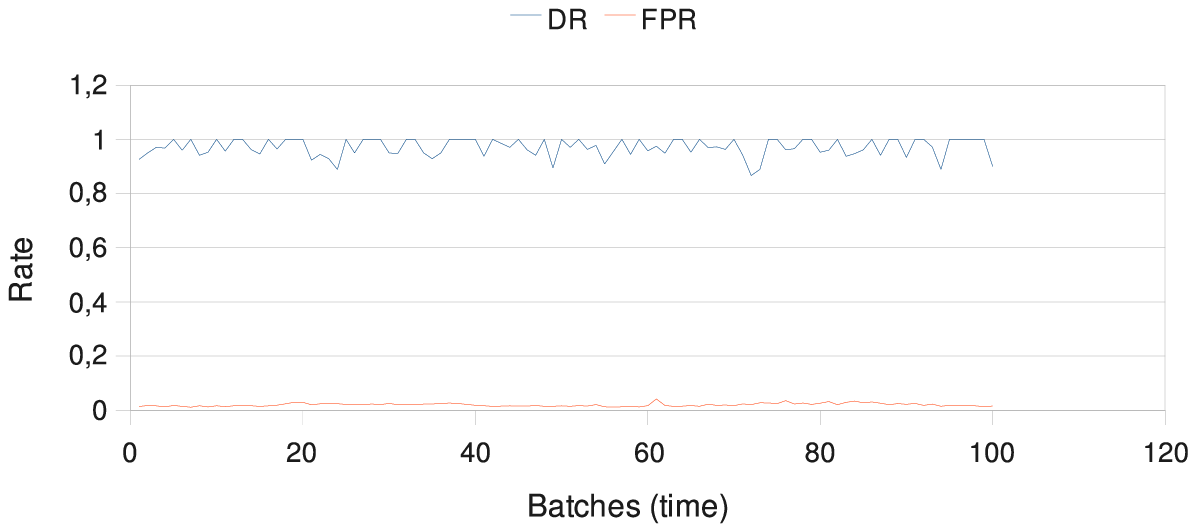} \\
 \includegraphics[width=0.49\textwidth]{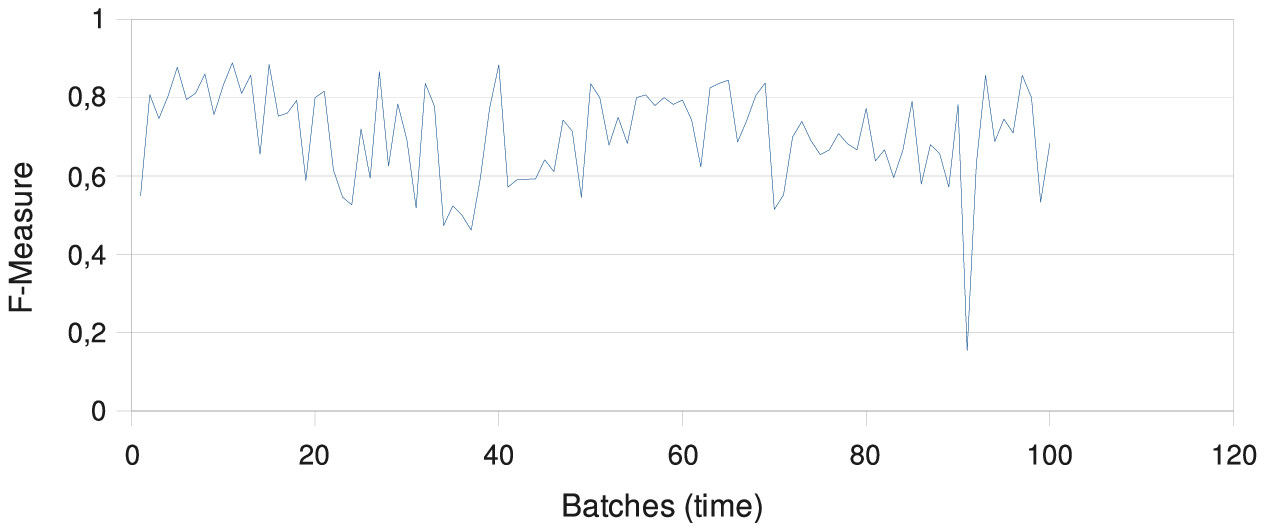}
\includegraphics[width=0.49\textwidth]{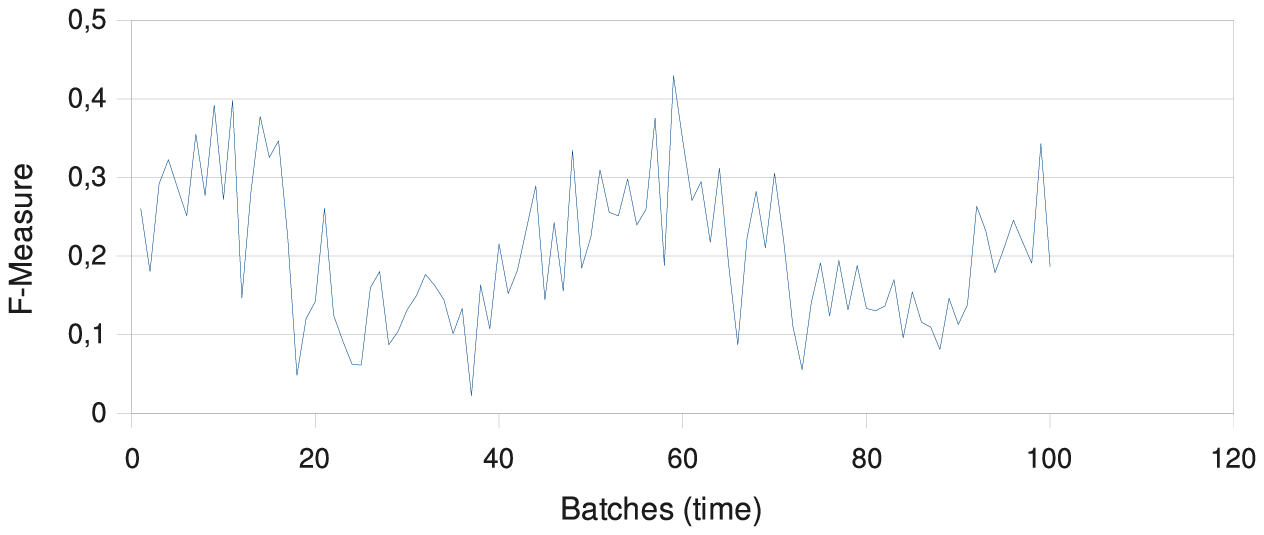}
 \caption{Evolution of the performance indicators over time. Figures in the left column: with adaptation, Figures in the left column: without adaptation. The first two figures (upper), give the evolution of the FP, FN and TP. Figures in the middle, give the evolution of the detection rate and the false positive rate. The last two Figures (lower), give the evolution of the F-Measure.}
\label{fig:Results}
\end{center}
\end{figure}

Figure \ref{fig:Results} shows the evolution of the performance indicators with and without adaptation. The Table \ref{tab:Results} shows the performance indicators computed with the entire log. With adaptation, 205565 adaptations occurred. The adaptation accuracy is more than 99\%. This means that there are only few cases in which the adaptation is faulty (\eg intrusive models are adapted with a normal request). The main part of the adaptation (202203 occurrences) consists in updating the normal model with a normal request.

\begin{table}
\center
\begin{tabular}{|p{2.5cm}|c|c|c|c|c|c|p{1.3cm}|}
\hline
 & \textbf{FP} & \textbf{FN} & \textbf{TP} & \textbf{TN} & \textbf{DR} & \textbf{FPR} & \textbf{F-Measure}\\ \hline
\textbf{With adaptation} & 2091 & 530 & 2018 & 997831 & 0.79 & 0.002 & 0.61 \\ \hline
\textbf{Without adaptation} & 21838 & 73 & 2461 & 978098 & 0.97 & 0.022 & 0.18 \\ \hline
\end{tabular}
\caption{Performance indicators computed with the diagnoses of the 1 million requests + 2534 intrusions}
\label{tab:Results}
\end{table}

The first two figures show that the adaptation reduces drastically the number of false positive diagnoses. The number of FP falls down at the very beginning. This means that the adaptation is quickly efficient, \ie the system is reactive. The number of FP does not increase thereafter. On the opposite, the number of FP stays around 200 per batch (without adaptation it is only about 50 FP per batch). Nonetheless, the true positive decreases a little while it is constant without adaptation. We can conclude that the adaptation makes our system more specific to intrusion: it efficiently reduces the false positive rate, but it reduces a little the true positives. The number of false negatives are low in the two cases.

The DR and FPR figures confirm that the detection performances with adaptation slowly decreases over time while it is constant without adaptation.
The average of detection rate is 80\%. Moreover, we see that the FPR is low, on average: 0.02 without adaptation and 0.002 with adaptation. Despite the only four diagnosers, these performances are quite good.

The F-Measure shows the global performances of a diagnoser, it takes into account both the sensitivity and the precision of the diagnoser to detect the intrusion. With adaptation, the F-Measure varies around 0.70($^{+}_{-}$0.14), while without adaptation, the F-Measure varies around 0.2($^{+}_{-}$0.09). Moreover, we can see that the global performances of the system with adaptation is relatively constant despite the decrease of the detection rate. In fact, the number of intrusion is very low, then the detection rate has only little influence on the global performances.

\espace

We explain the FPR difference with and without adaptation by the fact that our evaluation set of intrusions holds some intrusions that are not in the training set. With adaptation, the system discovers these initially unknown intrusions and the signatures have been enriched by this knowledge. On the opposite, without adaptation, the initially unknown intrusions stay unknown for the system and are not detected (false negatives).

The decrease of the true positive rate can be explained by an overlearning of normal models. In fact, a lot of adaptation of the normal model are performed and the models become overfitted. Consequently, some normal requests are less recognized over the time. In fact, diagnosis is a normalized distribution, if the normal mass is lower than before while the intrusion mass stays the same, the intrusion mass may become the biggest.

\espace

Based on these experiments, we can conclude that the adaptation improves reactively the global performances of the multi-diagnoser system and also maintains them at a high level. Moreover, our system discovers and detects dynamically new kinds of intrusion.

\section{Conclusion}

We have presented a system for self-adaptive intrusion detection from a stream of HTTP requests.
Our proposition associates a multi-diagnoser system and a meta-diagnosis process.
Each diagnoser agent constructs its diagnosis according to its own point of view of the system. Considering that the views are partially redundant, integrity constraints can be expressed on diagnoses. The meta-diagnosis process consists in using unsatisfied integrity constraints to trigger the adaptation of a subset of the diagnoser models.

The results of our experiments concluded that the multi-diagnoser architecture has good computing and performance results. The adaptation improves reactively the global performances of the multi-diagnoser system and also maintains them at a high level. Moreover, our system discovers and detects dynamically new kinds of intrusion. Nonetheless, we noticed that the number of true positives slowly decreases over time.

It is clear that more sophisticated methods can be used to locate the defective agents as for instance expert rules or any information on the source of the detected problem. Other heuristic or informed methods as well as model-based diagnosis methods (hitting-sets, prime implicants, etc.) could also be adapted to achieve this task.

\espace

We presented a general framework for adaptive intrusion detection system and its first application in order to prove the validity of our proposal. A first perspective will be to propose new diagnosers and new hypothesis to construct alternative diagnosis graph. For instance, it may be based on models learnt on the long term \textit{vs} models learnt recently, it may compare the diagnosis from a two different web servers (with their own signatures) may be compared, etc. We only used four diagnosers and the computing performances show that several diagnosers may be added without analyzing time constraints.

A second perspective is to design a more complete IDS solution.
We presented a completely autonomous system: not any manual action is required. Nonetheless, in real situations, it is strongly recommended to use the administrator expertise. For example, administrator may be helpful to correct (not necessarily frequently) the overlearning of some diagnosers that decreases slowly the performances of true positives diagnosis.
In such a case, our system may be used 1) as a tool adapting itself reactively to short term evolutions of the web server environment and 2) as a tool supporting the administrator to adapt the signatures base on the long term.
To support the administrators, our framework based on learnable signatures or models opens an new research direction in which the system would adapt their models from both their self-adaptation and interactive suggestion or correction from the administrator.

\bibliographystyle{abbrv}
\bibliography{mainBiblio}

\end{document}